\documentclass[11pt,twoside]{article} %TS01-3 %TS01-3 %TS23-2
\usepackage{cspm-asp2006}
\usepackage{epsfig,graphicx,natbib,url}  %RR additions
\usepackage{lscape} %%\usepackage{hyperref}
\begin{document}
\setcounter{page}{171}

\markboth{Tsiropoula et al.}{Multi-wavelength Analysis of a Quiet Region}
\title{Multi-wavelength Analysis of a Quiet Solar Region}
\author{G. Tsiropoula$^{1}$,
        K. Tziotziou$^{1}$,
        J. Giannikakis$^{1}$,
        P. Young$^{2}$,
        U. Sch\"{u}hle$^{3}$
        and
        P. Heinzel$^4$}
\affil{$^{1}$Institute for Space Applications and Remote Sensing, Athens, Greece\\
       $^{2}$CCLRC Rutherford Appleton Laboratory, United Kingdom\\
       $^{3}$MPI f\"{u}r Sonnensystemforschung, Katlenburg-Lindau, Germany\\
       $^{4}$Astronomical Institute AS, Ond\v{r}ejov, Czech Republic}
%%%%%%%%%%%%%%%%%%%%%%%%%%%%%%%%%%%%%%%%%%%%%%%%%%%%%%%%%%%%%%%%%%%%%%%%%%%%
\begin{abstract}
  We present observations of a solar quiet region obtained by the
  ground-based Dutch Open Telescope (DOT), and by instruments on the
  spacecraft SOHO and TRACE.  The observations were obtained during a
  coordinated observing campaign on October 2005. The aim of this work
  is to present the rich diversity of fine-scale structures that are
  found at the network boundaries and their appearance in different
  instruments and different spectral lines that span the photosphere
  to the corona.  Detailed studies of these structures are crucial to
  understanding their dynamics in different solar layers, as well as
  the role such structures play in the mass balance and heating of the
  solar atmosphere.
\end{abstract}
%%%%%%%%%%%%%%%%%%%%%%%%%%%%%%%%%%%%%%%%%%%%%%%%%%%%%%%%%%%%%%%%%%%%%%%%%%%%

%%%%%%%%%%%%%%%%%%%%%%%%%%%%%%%%%%%%%%%%%%%%%%%%%%%%%%%%%%%%%%%%%%%%%%%%%%%%
\section{Introduction}
%%%%%%%%%%%%%%%%%%%%%%%%%%%%%%%%%%%%%%%%%%%%%%%%%%%%%%%%%%%%%%%%%%%%%%%%%%%%

In the quiet regions of the solar surface the magnetic field is mainly
concentrated at the boundaries of the network cells. Over the past
decade, apart from the well-known mottles and spicules, several other
structures residing at the network boundaries such as explosive
events, blinkers, network flares, upflow events have been mentioned in
the literature. However, their interpretation, inter-relationship and
their relation to the underlying photospheric magnetic concentrations
remain ambiguous, because the same feature has a different appearance
when observed in different spectral lines and by different
instruments. For most of the events mentioned above magnetic
reconnection is suggested as the driving mechanism. This is not
surprising, since it is now well established from investigations of
high resolution magnetograms, that new bipolar elements emerge
continuously inside the cell interiors and are, subsequently, swept at
the network boundaries by the supergranular flow (\cite{tsir-wang};
\cite{tsir-schr}). Interactions of the magnetic fields have as a
result either the enhancement of the flux concentration in the case of
same polarities or its cancellation in the case of opposite
polarities. Observations support the idea that flux cancellation most
likely invokes magnetic reconnection. In this context, the study and
comprehension of the dynamical behaviour of the different fine-scale
structures is crucial to the understanding of the dynamics of the
solar atmosphere.

In this work we present observations of a solar quiet region and some
of the properties of several different structures appearing at the
network boundaries and observed in different wavelengths by the
different instruments involved in a coordinated campaign.

\begin{figure}
\centering
  \includegraphics[width=12cm]{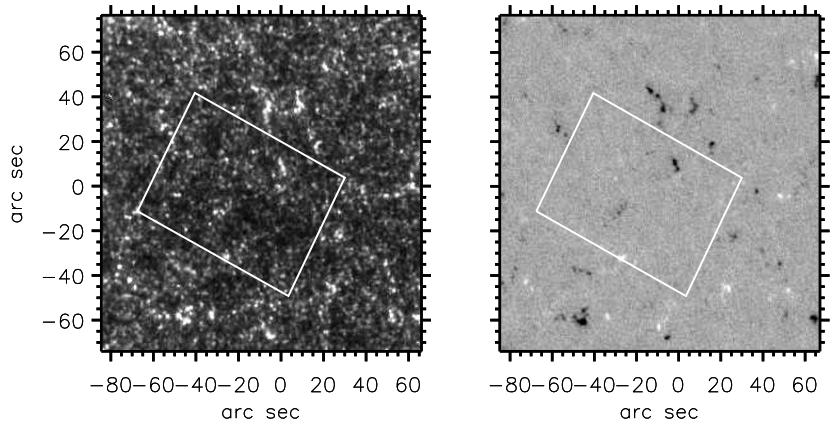}
  \caption[]{\label{fig:coalignment}
  {\em Left\/}: C\,IV TRACE image. {\em Right\/}: MDI magnetogram.
  The white rectangle inside the images marks the DOT's field-of-view.}
\end{figure}

%%%%%%%%%%%%%%%%%%%%%%%%%%%%%%%%%%%%%%%%%%%%%%%%%%%%%%%%%%%%%%%%%%%%%%%%%%%%
\section{Observations and Data Reduction}
%%%%%%%%%%%%%%%%%%%%%%%%%%%%%%%%%%%%%%%%%%%%%%%%%%%%%%%%%%%%%%%%%%%%%%%%%%%%

In October 2005 we ran a 12 days observational campaign. The aim
of that campaign was the collection of multi-wavelength
observations both from the ground and space that could be used for
the study of the dynamical behaviour of mottles/spicules and other
fine structures, observed in different layers of the solar
atmosphere.

Three ground-based telescopes were involved in that campaign: DOT
on La Palma,  THEMIS on Tenerife and SOLIS at Kitt Peak. From
space telescopes two spacecraft were involved: SOHO (with CDS,
SUMER, and MDI) and TRACE.

The analysed data were obtained on October 14 and consist of time
sequences of observations of a quiet region found at the solar disk
center recorded by different instruments. Sequences recorded by the
DOT were obtained between 10:15:43 -- 10:30:42 UT and consist of 26
speckle reconstructed images taken simultaneously at a cadence of
35\,s with a pixel size of 0.071\arcsec\ in 5 wavelengths along the
H$\alpha$\ line profile (i.e.\ at $-0.7$\,\AA, $-0.35$\,\AA, line
centre, 0.35\,\AA\ and 0.7\,\AA), in the G band with a 10\,\AA\ filter,
in the Ca\,II\,H line taken with a narrow band filter and in the blue
and red continuum. TRACE obtained high cadence filter images at
1550\,\AA, 1600\,\AA\ and 1700\,\AA. SUMER obtained raster scans and
sit-and-stare observations from 8:15 to 10:30 UT. CDS obtained
sit-and-stare observations from 6:44 to 10:46 UT and six 154\arcsec\
$\times$ 240\arcsec\ raster scans (each one having a duration of
30\,min) from 10:46 to 13:52 UT.  Both SOHO instruments (i.e., CDS and
SUMER) observed in several spectral lines spanning the upper solar
atmosphere. Using the standard software the raw measurements were
corrected for flat field, cosmic rays and other instrumental
effects. A single Gaussian with a linear background and Poisson
statistics were used for fitting each spectral line profile. MDI
obtained high cadence images at its high resolution mode.

Extensive work went into collecting, scaling and co-aligning the
various data sets to a common coordinate system (see
Fig.~\ref{fig:coalignment} showing the coalignment of TRACE, MDI, and DOT images).

\begin{figure}
\centering
  \includegraphics[width=12cm]{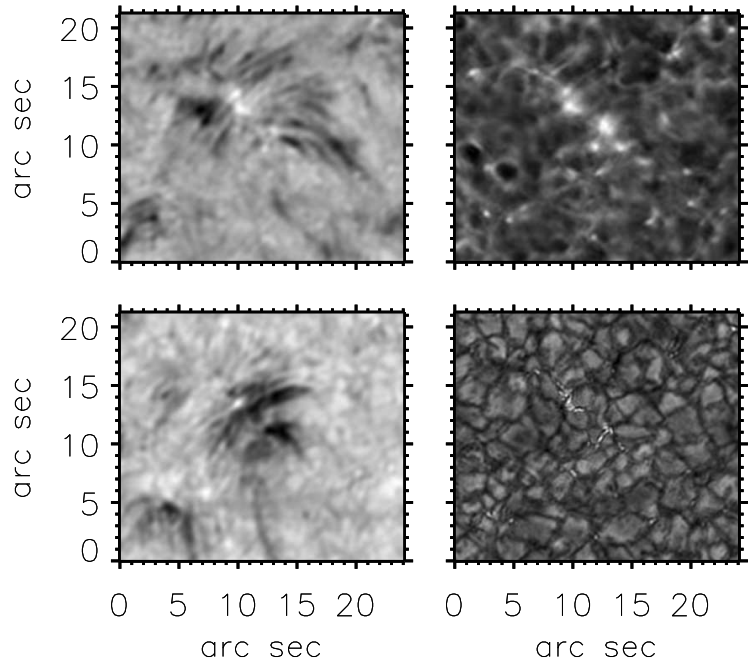}
  \caption[]{\label{fig:dotimage}
  DOT images of a rosette region. {\em Left\/}: H$\alpha$\,-0.7\,\AA\ {(\em first row\/)},
  H$\alpha$\,$+0.7$\,\AA\ {(\em second row\/)}. {\em Right\/}: Ca\,II\,H {(\em first row\/)},
  G band {(\em second row\/)}.}
\end{figure}

\section{Analysis and Results}
\subsection{DOT observations}
DOT's field-of-view (FOV) is 80\arcsec\ in the $X$ direction and
63\arcsec\ in the $Y$ direction. For the present study we selected a
smaller region which contains a rosette with several mottles
pointing to a common center (see Fig.~\ref{fig:dotimage}) and is
found at the middle upper part of the DOT's FOV. In the G-band
image (Fig.~\ref{fig:dotimage}, {\it right, second row}) isolated
bright points show up in the regions of strong magnetic field as
can be seen in the MDI image. These seem to be passively advected
with the general granular flow field in the intergranular lanes.
While we have not conducted an exhaustive study of bright point
lifetimes we find that bright points can be visible from some
minutes up to almost the entire length of the time series. In the
contemporal Ca\,II\,H images (Fig.~\ref{fig:dotimage}, {\it right,
first row} bright points are less sharp due to strong scattering
in this line and possibly due to increasing flux tube with height.
Reversed granulation caused by convection reversal is obvious in
this image.

In the H$\alpha-0.7$\,\AA\ (Fig.~\ref{fig:dotimage}, {\it left, first
row}) the dark streaks are part of the elongated H$\alpha$\ mottles seen
better at H$\alpha$\ line center. Some mottle endings appear extra dark
in the blue wing image through Doppler blueshift. Near the mottle
endings one can see bright points. These are sharper in the G-band
but stand out much clearer in the H$\alpha$\ wing. Thus H$\alpha$\ wing
represents a promising proxy magnetometer to locate and track
isolated intermittent magnetic elements. This is because the H$\alpha$\
wing has a strong photospheric contribution, as it is shown by
\citet{tsir-leen} who arrived to this conclusion by
using radiative transfer calculations and convective simulations.
In the H$\alpha+0.7$\,\AA\ (Fig.~\ref{fig:dotimage}, {\it left, second
row}) dark streaks around the rosette's center are signatures of
redshifts. Blueshifts in the outer endings and redshifts in the
inner endings of mottles provide evidence for the presence of
bi-directional flows along these structures.

\begin{figure}
\centering \includegraphics[width=11.5cm]{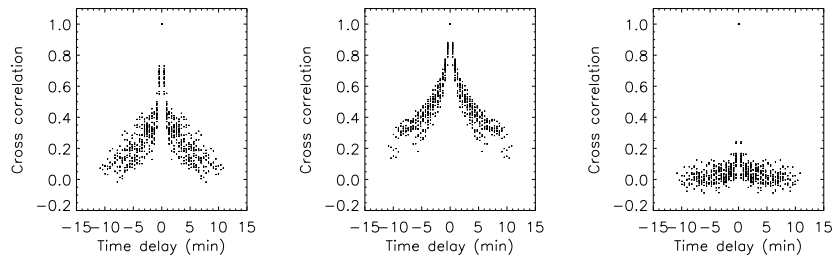}
  \caption[]{\label{fig:cross-cor} Cross correlation function vs time
  for {\em Left\/}: the intensity at H$\alpha-0.7$\,\AA, {\em
  Middle\/}: the intensity at H$\alpha$\ line center.  {\em Right\/}:
  the velocity at H$\alpha-0.7$\,\AA.}
\end{figure}

An important parameter for the study of the dynamics of mottles is
their velocity. For its determination, when filtergrams at two
wavelengths of equal intensity at the blue and the red side of the
line are available, a technique based on the subtraction of images
can be used. In this technique, by using the well known
representation of the line intensity profile and assuming a
Gaussian wavelength dependence of the optical thickness, we can
define the parameter DS:

\begin{equation}
DS = \frac{\Delta I}{\Sigma I-2I_{0\lambda}},
\end{equation}

\noindent where $\Delta I=I(-\Delta\lambda)-I(+\Delta\lambda)$,
$\Sigma I=I(-\Delta\lambda)+I(+\Delta\lambda)$ and
$I_0(\Delta\lambda)$ is the reference profile emitted by the
background. DS is called Doppler signal, has the same sign as the
velocity and can be used for a qualitative description of the velocity
field (for a description of the method see
\cite{tsir-tsiropoula}). When an optical depth less than one is
assumed then quantitative values of the velocity can be obtained from
the relation:

\begin{equation}
\upsilon =
\frac{{\Delta\lambda_{\rm D}}^2}{4\Delta\lambda}\frac{c}{\lambda}\ln\frac{1+DS}{1-DS},
\end{equation}

\noindent since in that case the velocity depends only on DS
(obtained from the observations) and the Doppler width,
$\Delta\lambda_{\rm D}$ (obtained from the literature). By using these
relations we have constructed 2-D intensity and velocity images
for the whole time series. We found out that it is difficult to
follow each one mottle for more than two or three frames and that
the general appearance of the region seems to change quite rapidly
with time. For a quantitative estimate of the temporal changes we
computed the value of the cross correlation (CC) function over the
2-D FOV both for the intensity and velocity. Figure~\ref{fig:cross-cor} shows the intensity CC curve at H$\alpha-0.7$\,\AA\
{\it (left)} and H$\alpha$\ line center {\it (middle)} and the velocity
CC curve at H$\alpha$\,$\pm$\,0.7\,\AA\ {\it (right)}. The decay of the CC
curve is a measure of the lifetime of the structures. The
e-folding time for the left curve is found equal to 2\,min, the
middle curve equal to 5\,min and the right curve is of the order
of the cadence.

\begin{figure}
\centering
  \includegraphics[width=4.8cm]{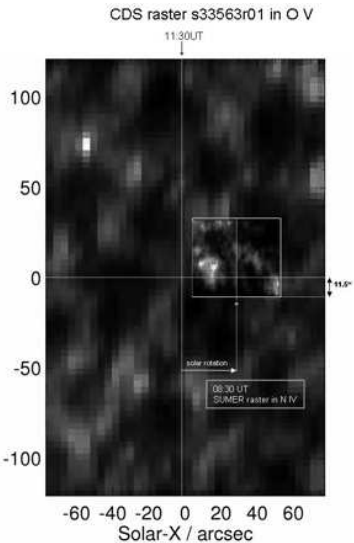}
  \caption[]{\label{fig:cds-sumer}
  CDS raster image obtained at the O\,V 629.7\,\AA\
  line with an overplotted SUMER image at Ne\,VIII 770\,\AA. }
\end{figure}

\subsection{CDS and SUMER observations}
In Fig.~\ref{fig:cds-sumer}\ we show the CDS raster image obtained
in the O\,V  629.7\,\AA\ line with the SUMER image at Ne\,VIII
770\,\AA\ overplotted. Although there is a time difference of
~3\,hours between the two images the network is constant enough to
allowed the coalignement of the two images. In CDS intensity maps
several brightenings are observed which are called blinkers
(\cite{tsir-harrison}). These events are best observed
in transition region lines and show an intensity increase of 60 -
80{\%}. Most of them have a repetitive character and reappear at
the same position several times.

\begin{figure}
\centering
  \includegraphics[width=9cm]{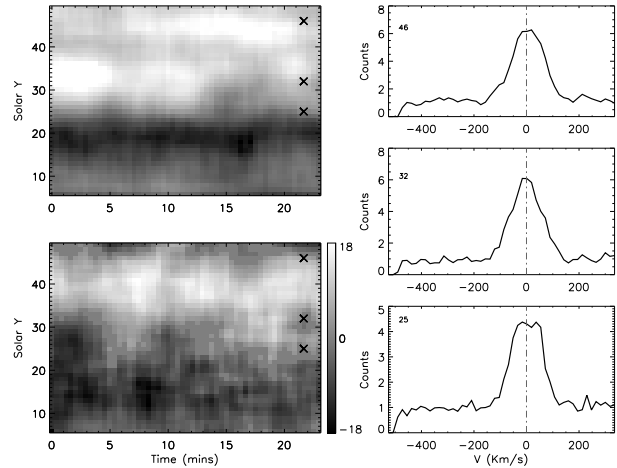}
  \caption[]{\label{fig:sumer}
  SUMER sit-and-stare observations in the Ne\,VIII line.
  {\em Left\/}: intensities ({\it up}), Doppler velocities ({\it
  bottom}) (network boundaries are bright in the intensity image).
  {\em Right\/}: non-Gaussian profiles in the positions
  marked by ``x'' inside the images in the left.
  }
\end{figure}

In Fig.~\ref{fig:sumer} ({\it left, up)} we show an integrated
(over the spectral line with the background included) intensity
image in the Ne VIII 770\,\AA\ line produced by sit-and-stare
observations of a network region. The image is produced by binning
over 6 spectra in order to improve the signal-to-noise ratio. The
Doppler shift map was derived by applying a single Gaussian
fitting (Fig.~\ref{fig:sumer}, {\it left, bottom}). The Doppler
shift map and the spectral line profiles were visually inspected
for any non-Gaussian profiles with enhancements in both the blue
and red wings that are the main characteristics of the presence of
bi-directional jets. Large numbers of such profiles were found at
the network boundaries (Fig.~\ref{fig:sumer}, {\it right}).

\section{Conclusions}
In this work we present observations of a quiet solar region
obtained by different instruments in different spectral lines.
Network boundaries are found to be the locus of several structures
which have different appearances when observed by different
instruments e.g., blinkers (when observed with CDS), mottles (when
observed with DOT), jets (when observed with SUMER). Their
interrelationship is to be further explored.

Regarding flows no-clear pattern is found in blinkers, while
bi-directional flows are found in jets. In dark mottles downward
velocities are found at their footpoints and upwards velocities at
their upper parts and very fast changes in their appearance.

The network shows a remarkable constancy when observed in low
resolution images. However when seen in high resolution images
several fine structures are observed which change so fast that it
is very hard to follow.

%%%%%%%%%%%%%%%%%%%%%%%%%%%%%%%%%%%%%%%%%%%%%%%%%%%%%%%%%%%%%%%%%%%%%%%%%%%%
\acknowledgements K. Tziotziou acknowledges support by Marie Curie
European Reintegration Grant MERG-CT-2004-021626. This work
has been partly supported by a Greek-Czech programme of
cooperation.

%RR all wrong bloody hell


\begin{thebibliography}{}

\bibitem[Harrison et al.(1999)]{tsir-harrison}
Harrison, A., Lang, J., Brooks, D.H., \& Innes, D.E. 1999, \aap,
351, 1115

\bibitem[Leenaarts et al.(2006)]{tsir-leen}
Leenarts, J., Rutten, R.J., S\"{u}tterlin, P., Carlsson, M., \&
Uitenbroek H. 2006, \aap, 449, 1209

\bibitem[Schrijver et al.(1997)]{tsir-schr}
Schrijver, C.J., Title, A.M., Van Ballegooijen, A.A., Hagenaar,
H.J., \& Shine, R.A. 1997, \apj, 487, 424

\bibitem[Tsiropoula(2000)]{tsir-tsiropoula}
Tsiropoula, G. 2000, New Astronomy, 5, 1

\bibitem[Wang et al.(2006)]{tsir-wang}
Wang, H., Tang, F., Zirin, H., \& Wang, J. 1996, \solphys, 165,
223


\end{thebibliography}
\end{document}